\documentclass{INTERSPEECH2023}
\usepackage{multirow}
\usepackage{hyperref}
\usepackage{algorithm}
\usepackage{algorithmicx}
\usepackage{algpseudocode}
\usepackage{threeparttable}
\usepackage{makecell}
\usepackage[dvipsnames]{xcolor}
\usepackage{subfigure}
\usepackage{color} 
\usepackage{colortbl}
\usepackage{tablefootnote}
\definecolor{applegreen}{rgb}{0.0, 0.5, 0.0}
\usepackage{float}
\interspeechcameraready

\title{BAT: Boundary aware transducer for memory-efficient and low-latency ASR}
\name{Keyu An, Xian Shi, Shiliang Zhang}
\address{Speech Lab of DAMO Academy, Alibaba Group, China}
\email{\{ankeyu.aky, shixian.shi, sly.zsl\}@alibaba-inc.com}

\begin{document}

\maketitle
 
\begin{abstract}
Recently, recurrent neural network transducer (RNN-T) gains increasing popularity due to its natural streaming capability as well as superior performance. Nevertheless, RNN-T training requires large time and computation resources as RNN-T loss calculation is slow and consumes a lot of memory. Another limitation of RNN-T is that it tends to access more contexts for better performance, thus leading to higher emission latency in streaming ASR. In this paper we propose boundary-aware transducer (BAT) for memory-efficient and low-latency ASR. In BAT, the lattice for RNN-T loss computation is reduced to a restricted region selected by the alignment from continuous integrate-and-fire (CIF), which is jointly optimized with the RNN-T model. Extensive experiments demonstrate that compared to RNN-T, BAT reduces time and memory consumption significantly in training, and achieves good CER-latency trade-offs in inference for streaming ASR.
\end{abstract}

\section{Introduction}
Recently, the recurrent neural network transducer (RNN-T)~\cite{rnn-t} has emerged as a promising end-to-end ASR framework due to its competitive performance and streaming-friendly nature. RNN-T models the acoustic and language features jointly, which eliminates the drawbacks in the output-independent CTC model~\cite{ctc}. 
Nevertheless, this appealing feature comes at the cost of high memory and computation consumption during training. Specifically, the RNN-T loss calculates on a 4-D lattice of shape (N, T, U, V), where N is the batch size, T is the output length of the acoustic encoder, U is the output length of the prediction network, and V is the vocabulary size. The large memory requirements limit the RNN-T training over large batches and hence slow down the training speed, especially for languages like Mandarin, where a large vocabulary set is adopted. To overcome it, frame downsampling or skipping~\cite{Wang2022AcceleratingRT} (reducing T), restrictions on the RNN-T lattice (reducing T or U)~\cite{kuang2022pruned,ar-rnnt}, sampled softmax (reducing V)~\cite{Sampled}, encoder and prediction output combination (reducing memory waste caused by padding)~\cite{li2019improving}, function merging (reducing memory waste caused by calculating and storing intermediate variables)~\cite{li2019improving}, and more efficient training pipeline (reducing training epochs)~\cite{Efficient-Training} have been studied in the previous work. In these works, the main challenge is to reduce memory consumption without degradation of accuracy.

Another important issue for RNN-T is the emission latency. RNN-T model optimized with unregularized loss tend to use more (future) context to produce better predictions, which causes significant emission delays~\cite{fastemit} (the difference between the user speaking and the model prediction). To address it, sequence-level emission regularization~\cite{fastemit} and token-level restrictions to enforce prediction within a reasonable time window~\cite{ar-rnnt, EmittingWT} were adopted in the previous work. Despite the progress, it is still challenging to obtain good WER-latency trade-offs.

\begin{figure}[!t]
  \centering
  \includegraphics[width=0.75\linewidth]{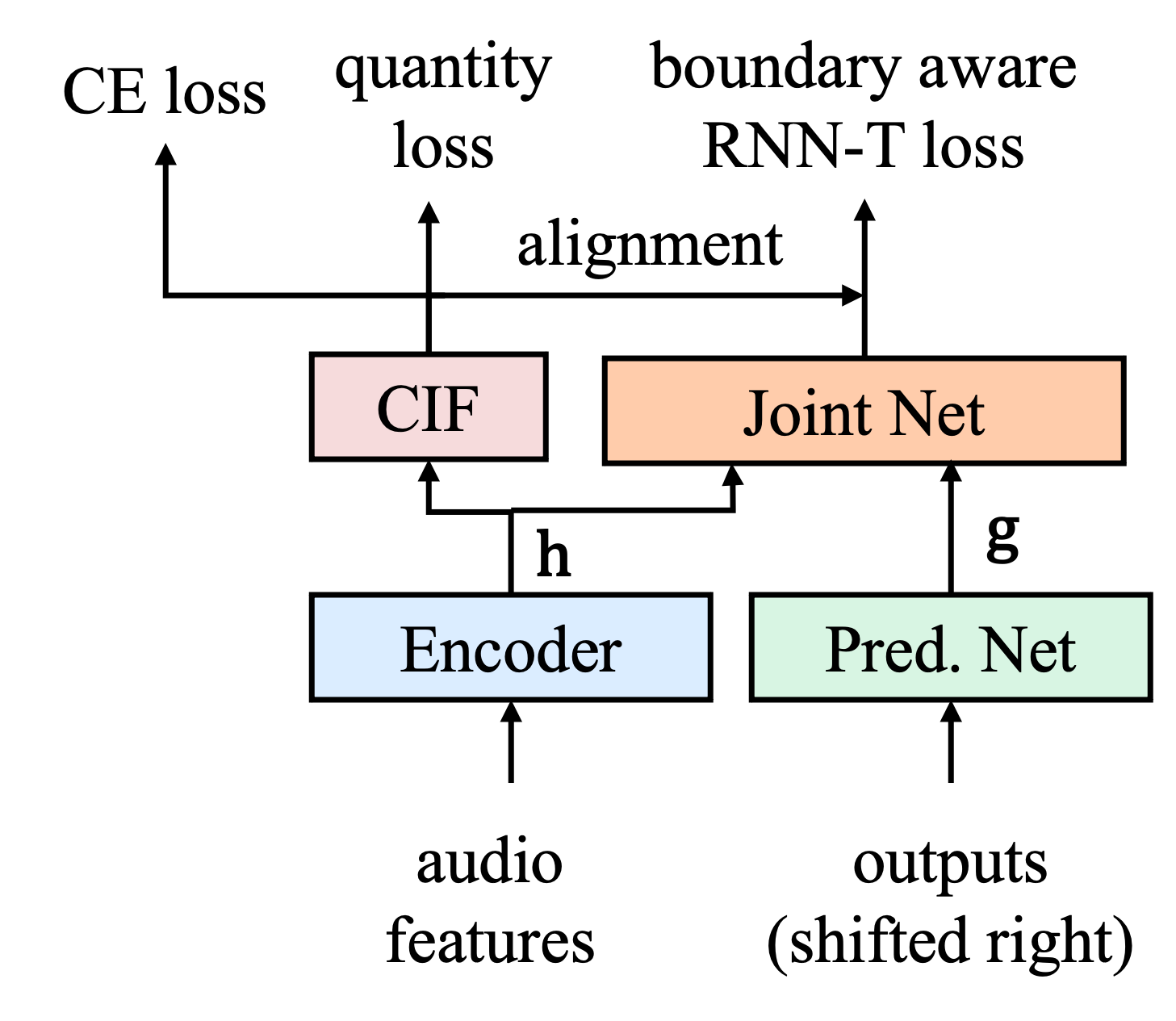}
  \caption{Boundary-aware training of RNN-T with the CIF alignment.}
  \label{fig:bat}
  %\vspace{-0.25cm}
\end{figure}

In this paper, we propose boundary-aware transducer (BAT) for memory-efficient, low-latency ASR, as illustrated in Fig.~\ref{fig:bat}. In BAT, the lattice for RNN-T loss calculation is reduced to a restricted region selected by the continuous integrate-and-fire (CIF) alignment~\cite{cif}. Thus, we only need to consider the forward variables $\alpha (t,u)$ and backward variables $\beta (t,u)$  within the limited region, which greatly reduces memory consumption in training. Moreover, the restricted alignment prevents the model from using unlimited context to produce the prediction, which leads to faster token emission for streaming inference. The CIF module is lightweight and fast, and thus can be jointly optimized with RNN-T from scratch, without great additional computation overhead. Results on the public AISHELL-1 datasets and non-public large-scale in-house data demonstrate that BAT achieves significant memory and latency reduction while maintaining recognition accuracy.

The paper is organized as follows. 
Section 2 outlines related work. Section 3 details the method of the boundary-aware transducer. Experiments are shown in Section 4. Section 5 discusses the limitations of the method and future works. Section 6 gives the conclusion.
\section{Related work}
\emph{Reducing the memory usage} for RNN-T training by performing the forward-backward calculation on a reduced lattice has been explored in several previous works. Alignment Restricted RNN-T (Ar-RNN-T)~\cite{ar-rnnt} proved that it's possible to restrict regions for each token using a pre-trained hybrid ASR model, which leads to less memory usage. In the recently proposed pruned RNN-T~\cite{kuang2022pruned}, the pruned boundary is generated by firstly computing the forward-backward algorithm on a "trivial" joint network, and then selecting the (t, u) pairs which contribute most to the final loss based on the output of the "trivial" joint network. 

For \emph{reducing the emission latency}, using a pre-generated alignment to constrain or guide the RNN-T training is also found to be beneficial. In these works, an external low-latency acoustic model (typically a hybrid model) is required to provide the frame-wise alignment. After that, the emission latency is penalized by masking out the alignment paths exceeding the predetermined threshold delay on the RNN-T lattice~\cite{ar-rnnt, EmittingWT}, or using an auxiliary loss~\cite{inaguma2020minimum}. Recently, sequence-level emission regularization methods propose to reduce latency by modifying the RNN-T loss to find paths tending to predict vocabulary tokens instead of blanks~\cite{fastemit, Delay-penalized}, without an external alignment. Nevertheless, the path may not be optimal for ASR due to a lack of alignment information, which can degrade the ASR accuracy severely~\cite{kim2021reducing, Delay-penalized}.

Inspired by the previous studies,  boundary-aware transducer tries to achieve memory and latency reduction at the same time, based on the alignment information. The major differences from previous works are 1) in BAT, the alignment information is generated on-the-fly using a lightweight and fast CIF module, which is jointly optimized with the RNN-T, thus not requiring a pre-trained (hybrid) model to obtain the token alignment. 2) the alignment generated by CIF is monotonic and continuous. Thus, there is no need to apply monotonic and continuity constraints to the token boundary, as pruned RNN-T did. 

\section{Method}
\subsection{RNN-T loss}
In the training stage, the RNN Transducer (RNN-T) model~\cite{rnn-t} aims to maximize the log-probability of a conditional distribution over the target token sequences ${\bf y} = (y_1, y_2, ..., y_U) \in \mathcal{Y}$ given the input sequence ${\bf x}=(x_1, x_2, ..., x_T)$:
$$\mathcal{L} = -{\rm log} {\rm Pr}({\bf y} | {\bf x}) =  -{\rm log} \sum_{{\bf a}\in \mathcal{B}^{-1}(y)} {\rm Pr}(\bf{a}| {\bf x}) $$
where ${\bf a} = (a_1, a_2, ..., a_{T+U}) \in \mathcal{Y} \cup \{\phi\}$ is the blank label $\phi$ augmented alignment sequence, and the mapping $\mathcal{B}$ is defined by removing $\phi$ in the input sequence. 

${\rm Pr}(\bf{a} | {\bf x})$ is further factorized as
\begin{equation}
%\label{eq:rnnt}
  \begin{aligned}
& {\rm Pr}({\bf a} | {\bf x}) =  \sum_{i=1}^{T+U} {\rm Pr}(a_i|h_{t_i}, g_{u_i})
  \end{aligned}
%  %\vspace{-1mm}
\end{equation}
where ${\bf h} = (h_1, h_2, ..., h_T) = {\rm Enc({\bf x})}$ is the high-level representation produced by the encoder, and $g_u$ is the prediction vector computed by the prediction network,
\begin{equation}
%\label{eq:pn}
g_u = {\rm PredictNet}({\bf y}_{[0: u-1]})
\end{equation}
, with the convention $y_0 = \phi$. The probability ${\rm  Pr}(\cdot | h_t, g_u)$ is typically implemented as the output of the joint network:
\begin{equation}
\label{eq:pr}
{\rm Pr}(\cdot | h_t, g_u) = {\rm softmax}[{\bf W}^{out} {\rm tanh}({\bf W}^{enc} h_t + {\bf W}^{pred} g_u + b)]
\end{equation}

To optimize the transducer objective, ${\rm Pr}({\bf y} | {\bf x})$ and gradients are calculated using the efficient forward-backward algorithm~\cite{rnn-t}:
$${\rm Pr}({\bf y} | {\bf x}) = \sum_{(t,u): t+u=n} \alpha (t,u) \beta (t,u), \forall n: 1 \leq n \leq T+U$$
$$\frac{\partial {\rm Pr}({\bf y} | {\bf x})}{\partial {\rm Pr}(a|h_t,g_u)} = \alpha (t,u)\left\{
\begin{aligned}
& \beta (t, u+1), {\rm if~} a = y_{u+1}, \\
& \beta (t+1, u), {\rm if~} a = \phi, \\
& 0, {\rm otherwise.}
\end{aligned}
\right.$$
where $\alpha (t,u)$ is the forward variable, defining the probability
of outputting $y_{[1:u]}$ during $h_{[1:t]}$, and $\beta (t, u)$ is the backward variable, defining the probability of outputting $y_{[u+1:U]}$ during $h_{[t:T]}$.  Denote
$$
y(t, u) =  {\rm Pr}(y_{u+1} | h_t, g_u) 
$$
$$
\phi (t, u) = {\rm Pr}(\phi | h_t, g_u) 
$$
Then the forward variables and backward variables can be calculated recursively using
\begin{equation}
\begin{aligned}
\alpha (t,u) &=  \alpha (t-1,u) \phi (t-1, u) + \alpha (t,u-1) y (t, u-1) \\
\beta (t,u) &= \beta (t+1,u) \phi(t, u)  +  \beta (t,u+1) y (t, u)
\end{aligned}
\end{equation}
It can be seen that RNN-T loss computation can consume a lot of time and memory because it has to compute $y(t, u)$ and $\phi(t, u)$ for all $0 \leq t \leq T$ and $0 \leq u \leq U$ on a lattice of shape (N, T, U, V). 

\subsection{Boundary aware training with CIF alignment}
\label{sec:cif}
\begin{figure}[!ht]
  \centering
  \includegraphics[width=\linewidth]{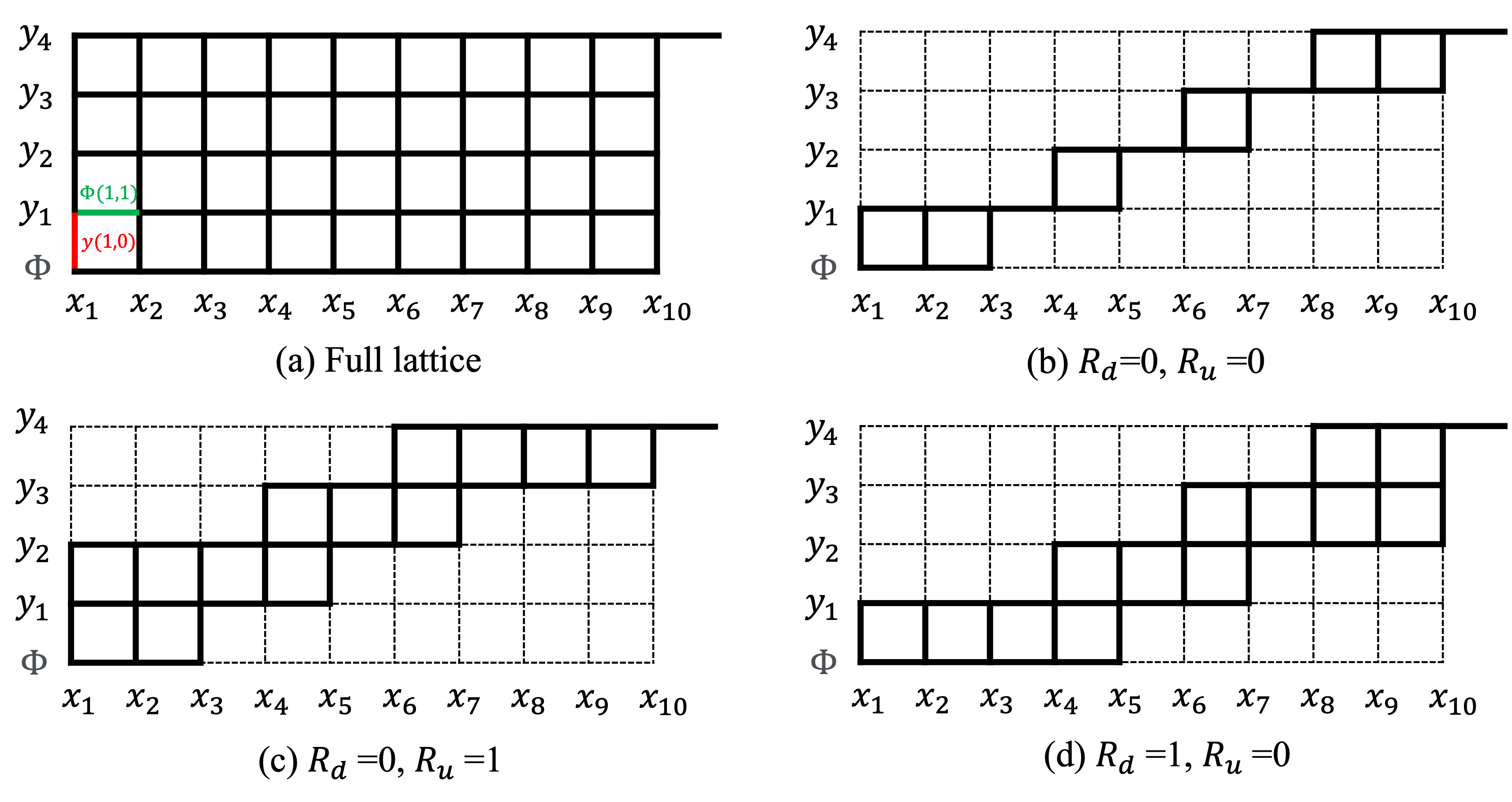}
  \caption{Output probability lattice defined by the joint network output. $(a)$ is the full lattice of standard RNN-T, and $(b)(c)(d)$ are restricted lattices with different $R_d$ and $R_u$, given the CIF alignment  $\mathcal{C}$  = $[1, 1, 1, 2, 2, 3, 3, 4, 4, 4]$. }
  \label{fig:ali}
  %\vspace{-0.25cm}
\end{figure}

In BAT, we use continuous integrate-and-fire (CIF)~\cite{cif}  to generate the monotonic alignment between the acoustic signals and token sequences and use it to restrict the paths being optimized during training, as illustrated in Fig.~\ref{fig:ali}. The pipeline of boundary-aware training is detailed below.

\textbf{CIF.} Given the encoder output ${\bf h} = (h_1, h_2, ..., h_T) $, CIF predicts the weights $ \bm{\omega} = ({\omega}_1, {\omega}_2, ..., {\omega}_T)$ using
\begin{equation}
\bm{\omega } = {\rm Sigmoid ({\rm Linear}({\rm Conv}({\bf h})))}
\end{equation}
Then, it forwardly accumulates the weights and integrates the encoder outputs until the accumulated weight reaches a given threshold $\beta_{\rm CIF}$, which means a token boundary is located.  It then instantly fires the integrated acoustic information for token prediction and updates the accumulated weights.  
In the training stage, the weights $ \bm{\omega}$ are scaled by $\frac{U}{\sum_{t=1}^{T} {\omega}_t}$ so that the predicted length of the token sequence is equal to the length of the target sequence and the model can be optimized using a simple cross-entropy (CE) loss $\mathcal{L}_{\rm CIF-CE}$, and quantity loss term 
\begin{equation}
\mathcal{L}_{\rm CIF-QUA} = \left| {\sum_{t=1}^{T} {\omega}_t} - U \right|
\end{equation}
is added to the total loss to encourage the model to predict the length of labels closer to the targets. We adopt a fast parallel implementation and a lightweight neural net configuration for the CIF module so the additional parameter and computation overhead introduced by CIF is insignificant.

\begin{figure}[]
  \centering
  \includegraphics[width=\linewidth]{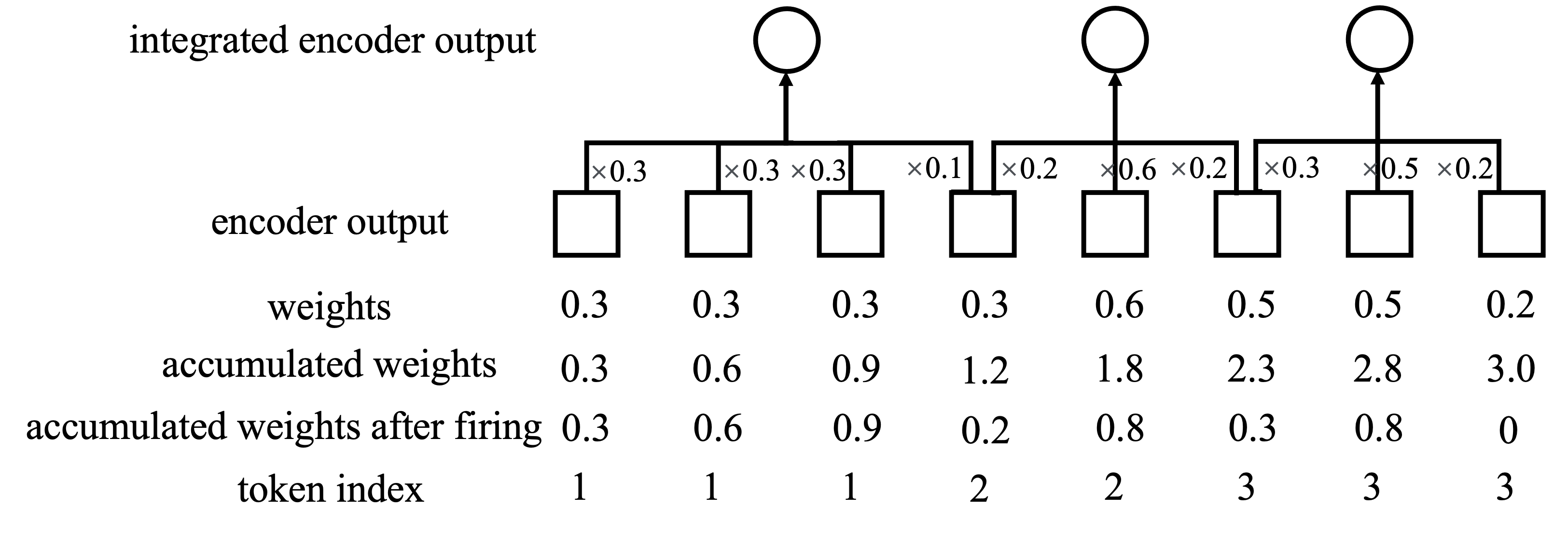}
  \caption{Token boundary generation with CIF. The weights have been scaled to match the target token number (3) and the threshold $\beta_{\rm CIF}$ is 1.}
  \label{fig:cif}
  %\vspace{-0.25cm}
\end{figure}

\textbf{Token boundary generation.} Given $\bm{\omega }$, we obtain the CIF alignment  $\bm{\mathcal{C}} = ({\mathcal{C}}_1, {\mathcal{C}}_2, ..., {\mathcal{C}}_T) $ sequence simply using 
\begin{equation}
\label{eq:cif}
\bm{\mathcal{C}} = {\rm ceil}({\rm cumsum}({\bm \omega}))
\end{equation}
, where $\mathcal{C}_t$ is defined as the index of the token to which $h_t$ assigns. 
${\rm ceil}()$ is the ceiling function and ${\rm cumsum}()$ is the cumulative sum operation, as illustrated in Fig.~\ref{fig:cif}. Note that the alignment given by eq.~\ref{eq:cif} is naturally monotonic ($\mathcal{C}_{t+1}$-$\mathcal{C}_t \geq$ 0) and continuous ($\mathcal{C}_{t+1}$-$\mathcal{C}_t \leq$ 1) so the ad hoc operations to make the pruning boundary valid in pruned RNN-T~\cite{kuang2022pruned} are avoided.

\textbf{Boundary aware training.} In boundary-aware training, we assume that $y(t, u)$ and $\phi(t, u)$ with non-zero gradients are located in a small neighboring region of CIF alignment. Thus, instead of considering $y(t, u)$ and $\phi(t, u)$ over all $U$ tokens at each time step $t$, we only compute $y(t, u)$ for 
$$\mathcal{C}_t-R_d \leq u \leq \mathcal{C}_t+R_u$$, and $\phi(t, u)$ for 
$$\mathcal{C}_t-R_d \leq u \leq \mathcal{C}_t+R_u+1$$. 
On the contrary, $y(t, u)$ for 
$$u \in \{ u \mid  u<\mathcal{C}_t - R_d~{\rm or}~u > \mathcal{C}_t + R_u \} $$
and $\phi(t, u)$ for 
$$u \in \{ u \mid u<\mathcal{C}_t - R_d~{\rm or}~u > \mathcal{C}_t + R_u + 1 \} $$
will be treated as 0.  
$R_d$ and $R_u$ are two hyper-parameters that control the ranges of the tokens that will be evaluated for time step $t$. 
Thus, the output lattice shape becomes (N, T, $R_d + R_u + 2$, V), which greatly reduces memory consumption. 
In the training stage, the CIF module is jointly optimized with the boundary-aware transducer (BAT) and the total training objective is defined as follows:
$$\mathcal{L} = \mathcal{L}_{\rm BAT} + \mathcal{L}_{\rm CIF-CE} + \mathcal{L}_{\rm CIF-QUA} $$

\section{Experiments}
\subsection{Experiment settings}
\label{sec:set}
We conduct our experiments on the openly available 170-hour Mandarin AISHELL-1~\cite{aishell1} dataset a 30000-hour in-house industrial Mandarin dataset.  The code and pre-trained model will be released upon publication of the paper.

On AISHELL-1, we use 80-dim filterbanks as input features. SpecAugment~\cite{specaug} and 3-fold speed perturbation are used for data augmentation. The encoder is a 12-layer conformer~\cite{Conformer}. The convolution kernel size is 31, and the number of attention heads, attention dimension, and feed-forward dimension are 8, 512, and 2048 respectively. We perform frame downsampling on 1) the input feature by a factor of 4 using stride convolution and 2) the output of the encoder by a factor of 2 to reduce the training and inference memory consumption. The prediction network is a 1-layer LSTM with 512 hidden units. The modeling units are 4233 Chinese characters. The CIF module consists of a 1D convolution layer with 512 channels and a linear layer. The total number of parameters is about 90M. For the streaming model, we adopt the causal convolution~\cite{Transformer-Transducer} and chunk attention based conformer. The convolution kernel size is 15 and the attention chunk size is 16. Other configurations are the same as the non-streaming model. We train the model from scratch for 100 epochs and average 10 checkpoints which perform best on the development set to obtain the final model. At the inference stage, we use beam search with a beam size 10, and no extra language model is used. 

On the 30000-hour in-house dataset, we stack the consecutive frames within a context window of 7 (3+1+3) to produce the 560-dimensional features and then perform 6$\times$ down-sampling on the input frames. The modeling units are 3445 Chinese characters. The model configurations are the same as the AISHELL-1. We train the model for 20 epochs and average 5 checkpoints which perform best on the development set.
\subsection{Latency metrics}
We use two metrics to characterize the latency for streaming RNN-T and streaming BAT.

\textbf{Average last token Emitting Time, avg ET.} The average time when the last token is emitted for all utterances in the test set.

\textbf{Partial Recognition (PR) Latency.}  The difference of the time (1) when the last token is emitted and (2) when a user finishes speaking, which is estimated by the alignment from the conventional model. Following~\cite{fastemit}, we report both 50-th (medium, PR50) and 90-th (PR90) percentile values of PR.
\subsection{Results}
\begin{table}[!ht]
\caption{The non-streaming RNN-T and boundary-aware transducer (BAT) results on AISHELL-1.}
%\vspace{0.25cm}
\begin{threeparttable}
	\centering
	\scalebox{1}{
	\begin{tabular}{l|cc|cc}
	\toprule
     \textbf{model}  & \textbf{$R_d$} & \textbf{$R_u$} & \textbf{dev CER} & \textbf{test CER}  \\
    \midrule
    RNN-T & - & - & 4.86 & \textbf{5.22}  \\
    BAT & 1 & 1 & 5.05 & 5.45 \\
    BAT & 2 & 2 & 4.86 & 5.32  \\  
    BAT & 3 & 3 & \textbf{4.82} & 5.28\\  
    \bottomrule
\end{tabular}}
\end{threeparttable}
\label{rnnt_results}
%\vspace{-0.5cm}
\end{table}
\begin{table*}[!h]
\caption{The streaming RNN-T and boundary-aware transducer (BAT) results on AISHELL-1. ET is the last token emitting time. PR is Partial Recognition Latency. FE is short for FastEmit. $\lambda$ is the FastEmit hyperparameter.}
%\vspace{0.25cm}
\begin{threeparttable}
	\centering
	\scalebox{1}{
	\begin{tabular}{c | lcc|cc|ccc}
	\toprule
     \textbf{Exp ID} & \textbf{model}  & \textbf{$R_d$} & \textbf{$R_u$} & \textbf{dev CER} & \textbf{test CER} & \textbf{avg ET (ms)} &  \textbf{PR50 (ms)} & \textbf{PR90 (ms)} \\
    \midrule
    1 & RNN-T &  - & - & 5.80 & 6.96 & 4633 &  100 & 230 \\
    2 & RNN-T + FE, $\lambda=0.002$ &   - & - & 6.11 & 7.44 & 4517 {\scriptsize \color{applegreen}~(-116)} & -20 {\scriptsize \color{applegreen}~(-120)}  & 120 {\scriptsize \color{applegreen}~(-110)} \\
    3 & RNN-T + FE, $\lambda=0.004$ &   - & - & 6.04 & 7.67 & 4416 {\scriptsize \color{applegreen}~(-217)} & -110 {\scriptsize \color{applegreen}~(-210)}  & 30 {\scriptsize \color{applegreen}~(-200)} \\
    4 & BAT &  1 & 1 & 5.89 & 7.63 & 4483 {\scriptsize \color{applegreen}~(-150)}& -10 {\scriptsize \color{applegreen}~(-110)} & 120 {\scriptsize \color{applegreen}~(-110)}\\  
    5 & BAT &  2 & 2 & 5.88 & 7.35 & 4523 {\scriptsize\color{applegreen}~(-110)}  &  10 {\scriptsize\color{applegreen}~(-90)} & 140 {\scriptsize\color{applegreen}~(-90)} \\    
    \bottomrule
\end{tabular}}
\end{threeparttable}
\label{rnnt_results_streaming}
%\vspace{-0.5cm}
\end{table*}
For the non-streaming model (Table~\ref{rnnt_results}), BAT and RNN-T perform comparably in accuracy, especially when $R_d$ and $R_u$ are relatively large ($R_d$=3 and $R_u$=3). A memory-CER trade-off is observed as smaller $R_d$ and $R_u$ lead to less memory usage but higher CERs.

For the streaming model (Table~\ref{rnnt_results_streaming}), we report results for BAT and RNN-T with and without FastEmit. It is shown that 1) similar to FastEmit~\cite{kim2021reducing, Delay-penalized}, boundary-aware training lead to accuracy degradation compared to the baseline RNN-T, 2) BAT achieves comparable CER-latency trade-offs to FastEmit, while BAT is more memory-efficient in training.
%We find that 1) while both FastEmit and boundary-aware training lead to accuracy degradation compared to the baseline RNN-T, BAT has lower CERs than RNN-T with FastEmit (Exp 3 - 4 vs. Exp 2). 2) there is a CER-latency trade-off as models with lower CERs tend to have higher latency (Exp 2 - 5). 
%the performance of BAT is inferior to RNN-T, presumably because 1) it's more difficult to locate the token boundary for CIF when the encoder works in a streaming fashion 2) in the streaming model, the region located by CIF is not aligned with the region with non-zero gradient on the RNN-T lattice. A CER-latency trade-off is also observed as smaller pruning ranges lead to lower latency but higher CERs.

\begin{figure}[!ht]
  \centering
  \includegraphics[width=0.8\linewidth]{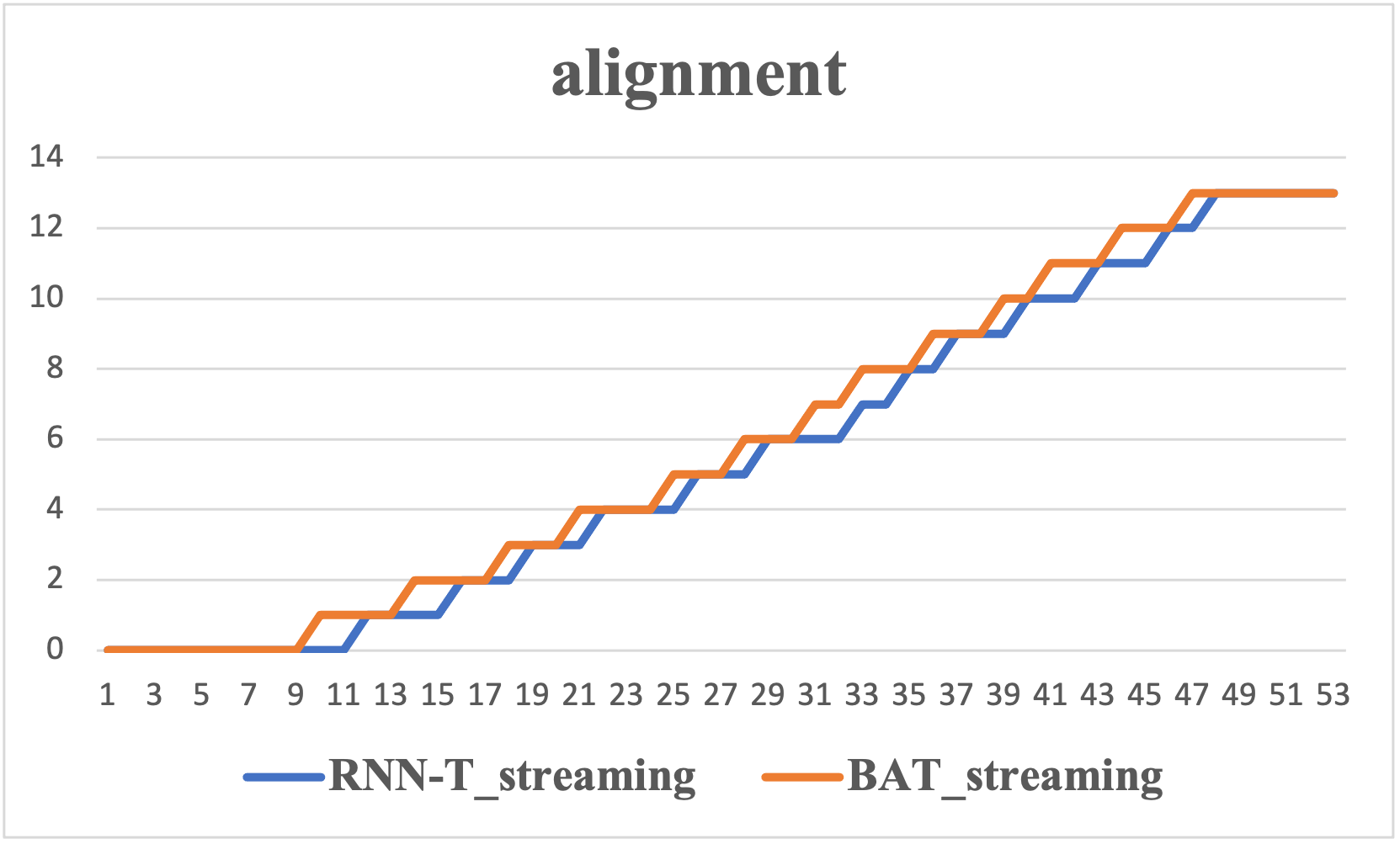}
  \caption{Alignment of BAT ($R_d$=2, $R_u$=2, colored in orange) and RNN-T (blue) for the streaming model. The Y-axis is the non-blank token emitted by the model, and the X-axis is the time step.}
  \label{fig:curve}
  %\vspace{-0.25cm}
\end{figure}

To better visualize the effectiveness of boundary-aware training in improving emission latency, we show the alignment for an utterance in the AISHELL-1 test set given by RNN-T and BAT in Fig.~\ref{fig:curve}. It is shown that boundary-aware training encourages a faster emission of tokens at inference.

The results of the in-house 30000-hour task (table \ref{rnnt_results_8k}) prove that the proposed method can be successfully applied to RNN-T training on large-scale data. 
\begin{table}[!ht]
\caption{The results of RNN-T and BAT ($R_d$=2, $R_u$=2) on the industrial 30000-hour task.}
%\vspace{0.25cm}
\begin{threeparttable}
	\centering
	\scalebox{1}{
	\begin{tabular}{l|c c}
	\toprule
     \textbf{model}  & \textbf{streaming} & \textbf{CER}    \\
    \midrule
    RNN-T & N & 8.08  \\
    BAT & N & 8.12 \\
    \midrule
    RNN-T & Y & 8.75  \\
    BAT & Y & 8.87 \\
    \bottomrule
\end{tabular}}
\end{threeparttable}
\label{rnnt_results_8k}
%\vspace{-0.5cm}
\end{table}

\subsection{Time and memory usage}
We benchmark the time and memory usage on the AISHELL-1 training data. The experiments are conducted on 1 Tesla V100 GPU. 
The model configuration is the same as the AISHELL-1 non-streaming model in Sec.~\ref{sec:set}. 
We sort utterances by the input feature lengths and set the number of frames (after padding) in a batch per GPU to 50k. 
Table~\ref{memory_usage} compares time and peak memory usages~\footnote{The memory allocating information is collected using the pytorch {max\_memory\_allocated} API.} for RNN-T loss calculation with warp-rnnt\tablefootnote{https://github.com/1ytic/warp-rnnt/tree/master/pytorch\_binding}, pruned RNN-T\tablefootnote{https://github.com/danpovey/fast\_rnnt} and BAT. We report stats for joint network and loss calculation as the encoder and prediction network calculations for the three implementations are the same. For pruned RNN-T, the time and memory statistics includes the trial joint network computation, and the number of indexes that will be evaluated for any time step  (S in the original paper) is set to 5 by default. For BAT, the time and memory statistics for CIF computation are included, and $R_d=R_u$=2. It is shown that both pruned RNN-T and BAT reduce time and memory consumption drastically. BAT outperforms pruned RNN-T, which indicates that pruning bounds generated by CIF are more efficient than that in pruned RNN-T. 
\begin{table}[!ht]
\caption{Time per batch (ms) and peak memory usage (GB) for RNN-T, pruned RNN-T and BAT ($R_d$=2, $R_u$=2).}
%\vspace{0.25cm}
\begin{threeparttable}
	\centering
	\scalebox{1}{
	\begin{tabular}{l| c c }
	\toprule
     \textbf{model} & \textbf{time (ms)} & \textbf{peak mem usage (GB)} \\
    \midrule
    warp-rnnt & 230 & 16.9 \\
    pruned RNN-T & 94 & 7.4  \\
    BAT & 85 & 6.4 \\
    \bottomrule
\end{tabular}}
\end{threeparttable}
\label{memory_usage}
%\vspace{-0.5cm}
\end{table}
\section{Limitations and future work}
In BAT, the token boundary information is only used in the training stage, and the inference of BAT is exactly the same as the standard RNN-T. In fact, the CIF alignment could also be used to guide and speed up the RNN-T inference (e.g. perform frame skipping~\cite{tian2021fsr} based on the CIF weights), which would be our future work.
\section{Conclusions}
In this paper we propose boundary-aware transducer (BAT) for memory-efficient and low-latency ASR. Different from previous works which utilize alignment references generated from an external pre-trained model, BAT can be end-to-end optimized, as the alignment for BAT is generated on the fly efficiently. Extensive experiments demonstrate that compared to RNN-T, BAT reduces time and memory consumption significantly in training, and achieves good CER-latency trade-offs in streaming inference.

\bibliographystyle{IEEEtran}
\bibliography{mybib}

\end{document}